\documentclass[twocolumn, times, tighten]{aastex61}
\usepackage{textcomp}
\usepackage{amsmath}
\usepackage{color}
\usepackage{amssymb}
\usepackage{hyperref}

\newcommand{\CIV}{C\,{\sc iv}}

\newcommand{\Hbeta}{H$\beta$}

\shorttitle{Sun et al.}

\begin{document}
\revised{\bf Draft: \today}
\title{A falling Corona model for the anomalous behavior of the broad emission lines in NGC 5548}

\author[0000-0002-0771-2153]{Mouyuan Sun}
\affiliation{CAS Key Laboratory for Research in Galaxies and Cosmology, 
Department of Astronomy, University of Science and Technology of China, Hefei 
230026, China; ericsun@ustc.edu.cn; xuey@ustc.edu.cn}
\affiliation{School of Astronomy and Space Science, University of Science 
and Technology of China, Hefei 230026, China}

\author[0000-0002-1935-8104]{Yongquan Xue}
\affiliation{CAS Key Laboratory for Research in Galaxies and Cosmology, 
Department of Astronomy, University of Science and Technology of China, Hefei 
230026, China; ericsun@ustc.edu.cn; xuey@ustc.edu.cn}
\affiliation{School of Astronomy and Space Science, University of Science 
and Technology of China, Hefei 230026, China}

\author[0000-0002-4223-2198]{Zhenyi Cai}
\affiliation{CAS Key Laboratory for Research in Galaxies and Cosmology, 
Department of Astronomy, University of Science and Technology of China, Hefei 
230026, China; ericsun@ustc.edu.cn; xuey@ustc.edu.cn}
\affiliation{School of Astronomy and Space Science, University of Science 
and Technology of China, Hefei 230026, China}

\author[0000-0001-8416-7059]{Hengxiao Guo}
\affiliation{National Center for Supercomputing Applications, University of Illinois at Urbana-Champaign, 
605 East Springfield Avenue, Champaign, IL 61820, USA}
\affiliation{Department of Astronomy, University of Illinois at Urbana-Champaign, Urbana, IL 61801, USA}

\begin{abstract}
NGC 5548 has been intensively monitored by the AGN Space Telescope and Optical Reverberation 
Mapping collaboration. Approximately after half of the light curves, the correlation between 
the broad emission lines and the lag-corrected ultraviolet continua becomes weak. 
This anomalous behavior is accompanied by an increase of soft X-ray 
emission. We propose a simple model to understand this anomalous behavior, i.e., the corona 
might fall down, thereby increasing the covering fraction of the inner disk. Therefore, X-ray 
and extreme ultraviolet emission suffer from spectral variations. 
The ultraviolet continua variations are driven by both X-ray and extreme ultraviolet 
variations. Consequently, the spectral variability induced by the falling corona would dilute 
the correlation between the broad emission lines and the ultraviolet continua. Our model can 
explain many additional observational facts, including the dependence of the anomalous behavior on 
velocity and ionization energy. We also show that the time lag and correlation 
between the X-ray and the ultraviolet variations change as NGC 5548 displays the anomalous 
behavior. The time lag is dramatically longer than the expectation from disk reprocessing 
if the anomalous behavior is properly excluded. During the anomalous state, the time lag approaches 
the light-travel timescale of disk reprocessing albeit with a much weaker correlation. We speculate 
that the time lag in the normal state is caused by reprocessing of the broad line region gas. 
As NGC 5548 enters the abnormal state, the contribution of the broad line 
region gas is smaller; the time lag reflects disk reprocessing. We also discuss alternative scenarios. 
\end{abstract}


\section{Introduction}
\label{sect:intro}
Variations of active galactic nucleus (AGN) emission often provide model-independent constraints 
on the size of the emission region. This approach is particularly effective in studies of the 
cross correlation between two light curves of AGN emission. For instance, the cross 
correlation between the ultraviolet (UV) to optical variations of continua and broad emission lines 
(BELs) can probe the distance between the broad emission line region (BLR) and the ionizing continua 
(which is the extreme-UV, i.e., EUV, emission) and reveal the structure of the BLR. Similarly, the 
cross correlation between any two light curves of AGN continua can constrain the size of the central 
engine, which is widely believed to be powered by accretion of materials onto a supermassive black 
hole (SMBH). 

Theoretically speaking, these correlations are causal in nature. The BLR is photoionized 
by the intense EUV photons and emits both high- (e.g., \CIV) and low-ionization (e.g., \Hbeta) 
BELs. As a result, the BELs vary in response to the ionizing continuum variations 
after light-travel time delays. This is the ``reverberation mapping'' of BELs \citep[RM; 
see][]{BM1982}. This technique, which usually requires time-resolved spectroscopy, has been 
applied to some AGNs that are diverse in various properties \citep{Kaspi2000, Kaspi2007, Peterson2002, 
Peterson2014, Bentz2010b, Denney2010, Grier2012, Grier2017b, Du2014, Bentz2015, Shen2016}. 
Empirical relations between SMBH mass ($M_{\rm{BH}}$) and the size and the velocity dispersion 
of the BLR have been established \citep[which are calibrated by the scaling relations between 
SMBHs and the physical properties of the bulges of their host, see, e.g.,][]{Onken2004}. These 
relations are referred as the single-epoch virial $M_{\rm{BH}}$ estimators \citep{Vestergaard2006}. 
Moreover, with high-quality RM data, it is possible to produce velocity-delay maps of BLR and 
constrain the kinematics of the BLR gas \citep[e.g.,][]{Denney2009, Bentz2010a, Grier2012, 
Grier2017a}. 

The inter-band correlations and time lags might be explained by the X-ray 
reprocessing scenario \citep[e.g.,][]{Krolik1991}. In the model, the highly variable X-ray 
emission can illuminate and heat the surface layer of the outer accretion disk \citep[for 
a review, see][]{Reynolds2003}. The surface layer then re-radiates the variable thermal 
emission that is at least partially responsible for the observed variations in UV, optical 
and near infrared (NIR) bands. Therefore, similar to that of the BLR case, the UV, optical 
and NIR emission 
follows the X-ray continuum variations after light-travel time delays. This model has been 
well developed and adopted to fit the multi-band light curves \citep[e.g.,][]{Edelson1996, 
Edelson2015, Edelson2017, Wanders1997, Collier1998, Sergeev2005, McHardy2014, McHardy2016, 
McHardy2017, Shappee2014, Fausnaugh2016, Starkey2016, Cackett2017, Starkey2017, Pal2018}. It is found 
that the detected inter-band time lags are inconsistent with the classical thin disk theory 
\citep[e.g.,][]{Fausnaugh2016, Starkey2017}. Moreover, the time lag between X-ray and UV 
emission is significantly longer than the light-travel timescale (e.g., \citealt{Edelson2017, 
McHardy2017}; but see \citealt{McHardy2016}). Other scenarios and characteristic 
timescales, e.g., the dynamical timescales, are proposed to explain the observations 
\citep[e.g.,][]{Cai2017, Gardner2017}. 

NGC 5548, a well-studied RM AGN, has been frequently monitored by the most intensive RM 
experiment to date, the AGN Space Telescope and Optical Reverberation Mapping (STORM) 
collaboration. During this RM campaign, NGC 5548 was observed with space \citep[including 
\textit{Swift}, the \textit{Hubble Space Telescope} (\textit{HST});][]{Derosa2015, Edelson2015} 
and ground-based telescopes \citep{Pei2017} with high cadence. The good time sampling, high 
signal-to-noise ratio (S/N), multiwavelength (including X-ray, UV, optical, and infrared) 
continua, and spectroscopic observations have guaranteed to explore the high degree of detail 
of the BLR \citep{Derosa2015, Pei2017}, the accretion disk \citep{Edelson2015, Fausnaugh2016}, 
and the corona \citep{Edelson2015}. 

As noted by \cite{Goad2016}, the variations of BELs in NGC 5548 underwent an interesting 
and unexpected anomalous behavior. Approximately after half of the observations, 
the BELs and the UV continua varied almost independently. During this ``abnormal'' state, 
the X-ray spectra underwent a significant evolution \citep{Mathur2017}. The anomalous BEL 
variations infringe the basic RM assumptions and require a physical explanation. In this 
work, we propose a simple falling corona model, aiming to explain this transient 
anomalous phenomenon. We also illustrate that the abnormal state has an effect on the time 
lag measurements between X-ray and UV variations. 

This paper is formatted as follows. In Section~\ref{sect:obsf}, we present the observational 
facts of the anomalous behavior in NGC 5548. In Section~\ref{sect:model}, we detail our 
simple model. We summarize our results in Section~\ref{sect:summ}. In this work, we adopt 
$M_{\mathrm{BH}}=5\times 10^7\ M_{\odot}$ for NGC 5548 \citep{Bentz2015}.

\section{Observational facts and our model}
\subsection{Observational facts}
\label{sect:obsf}
NGC 5548 was intensively monitored by the AGN STORM project. The time baseline of the campaign 
is $\sim 180$ days (i.e., between HJD=$2,456,690$ and HJD=$2,456,866$, or from 2014 February 1 
through 2014 July 27). During the first half of the AGN STORM campaign, BELs and UV emission vary 
coherently. However, the coherence disappears starting on THJD $\sim 6747$ days (THJD=HJD$-2,450,000$) 
even if the mean flux is similar to that of the first half of the light curve (see Figure~1 of 
\citealt{Goad2016} and Figure~7 of \citealt{Pei2017}). The observational constraints of the 
anomalous behavior in NGC 5548 are enumerated as follows \citep{Goad2016, Mathur2017, Pei2017}. 
\begin{itemize}
\item[1.] The response of the BELs to the UV continua behaves anomaly starting approximately 
midway through the campaign \citep{Goad2016, Pei2017}. To our surprise, the BELs and the UV 
continua are uncorrelated during this abnormal state. Meanwhile, the responsivity (i.e., the ratio 
of the changes of the BEL fluxes to the changes of the continuum) 
also decreases (i.e., the line emission is ``lost''). Such anomalous behaviors are found 
for both high- (e.g., CIV) and low-ionization (e.g., H$\beta$) emission lines. The BLR structure 
cannot change in such a short timescale; the anomalous behaviors of BELs are likely caused 
by a depletion of EUV photons. 

\item[2.] The fraction of emission line lost with respect to the UV fluxes, $f_{\mathrm{lost}}$, 
is defined as  $f_{\mathrm{lost}}=1-f_{\mathrm{obs}}/f_{\mathrm{rec}}$, where 
$f_{\mathrm{obs}}$ and $f_{\mathrm{rec}}$ are the observed and the expected \citep[which 
is reconstructed from the \textit{HST} $1367\ \rm{\AA}$ light curve;][]{Goad2016} BEL fluxes, 
respectively. It is found that $f_{\mathrm{lost}}$ is distinctly velocity dependent. $f_{\mathrm{lost}}$ 
increases from the line center to the wings \citep[see Figure~10 of][]{Pei2017}. 

\item[3.] The fraction of line emission lost for the high-ionization BELs is higher 
than that of the low-ionization BELs. For instance, while the fraction of CIV lost is 
$\sim 18\%$, the fraction of H$\beta$ lost is only $\sim 6\%$ \citep[see Table~6 of][]{Pei2017}.

\item[4.] During the abnormal state, the fraction of soft X-ray excess increases \citep{Mathur2017}. 

\item[5.] The abnormal state lasts for roughly $\sim 50$ days \citep{Goad2016, Pei2017}. 
NGC 5548 then returns to a normal state. 
\end{itemize}

\subsection{The model}
\label{sect:model}
\begin{figure}
\epsscale{1.8}
\plottwo{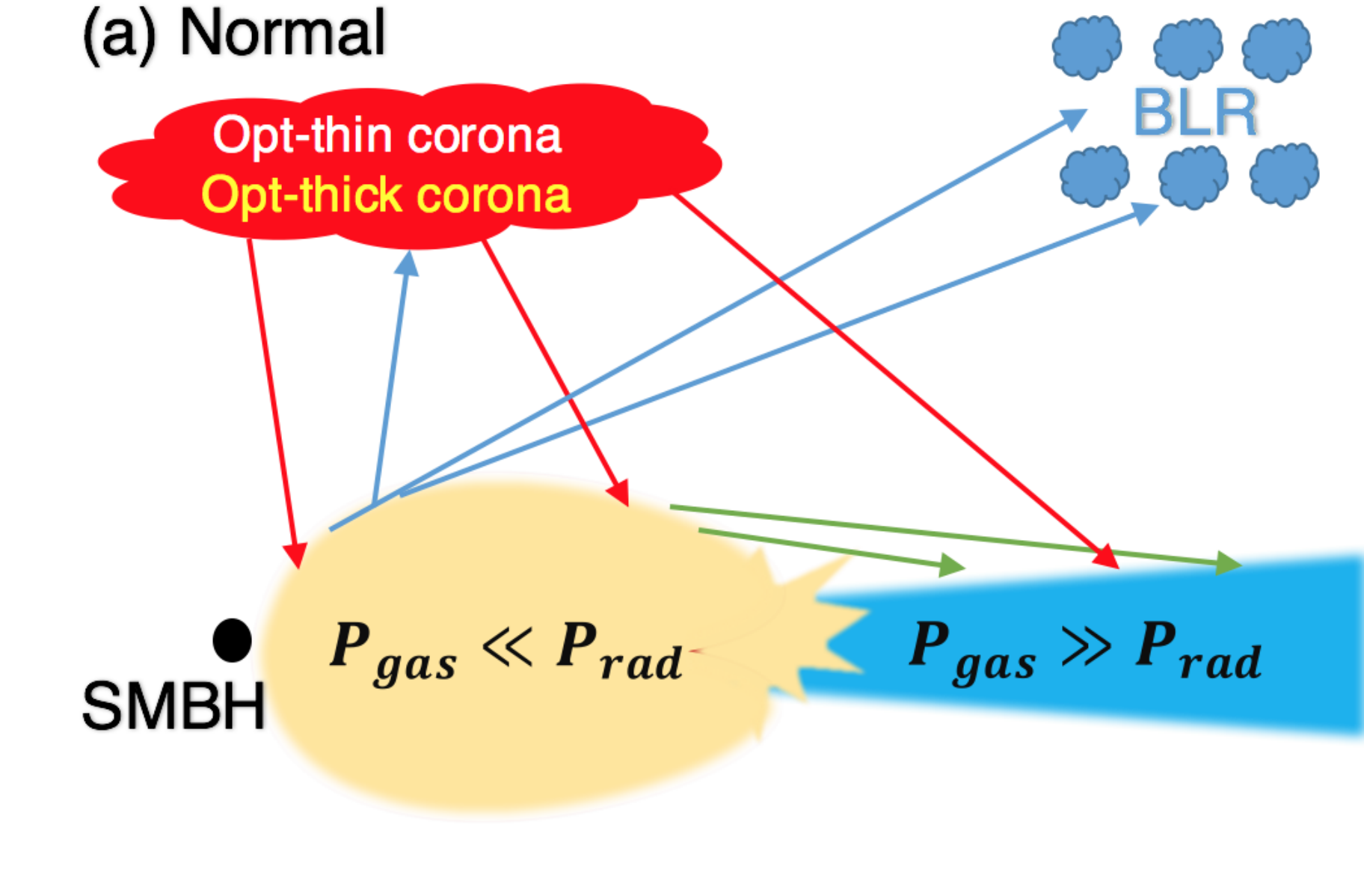}{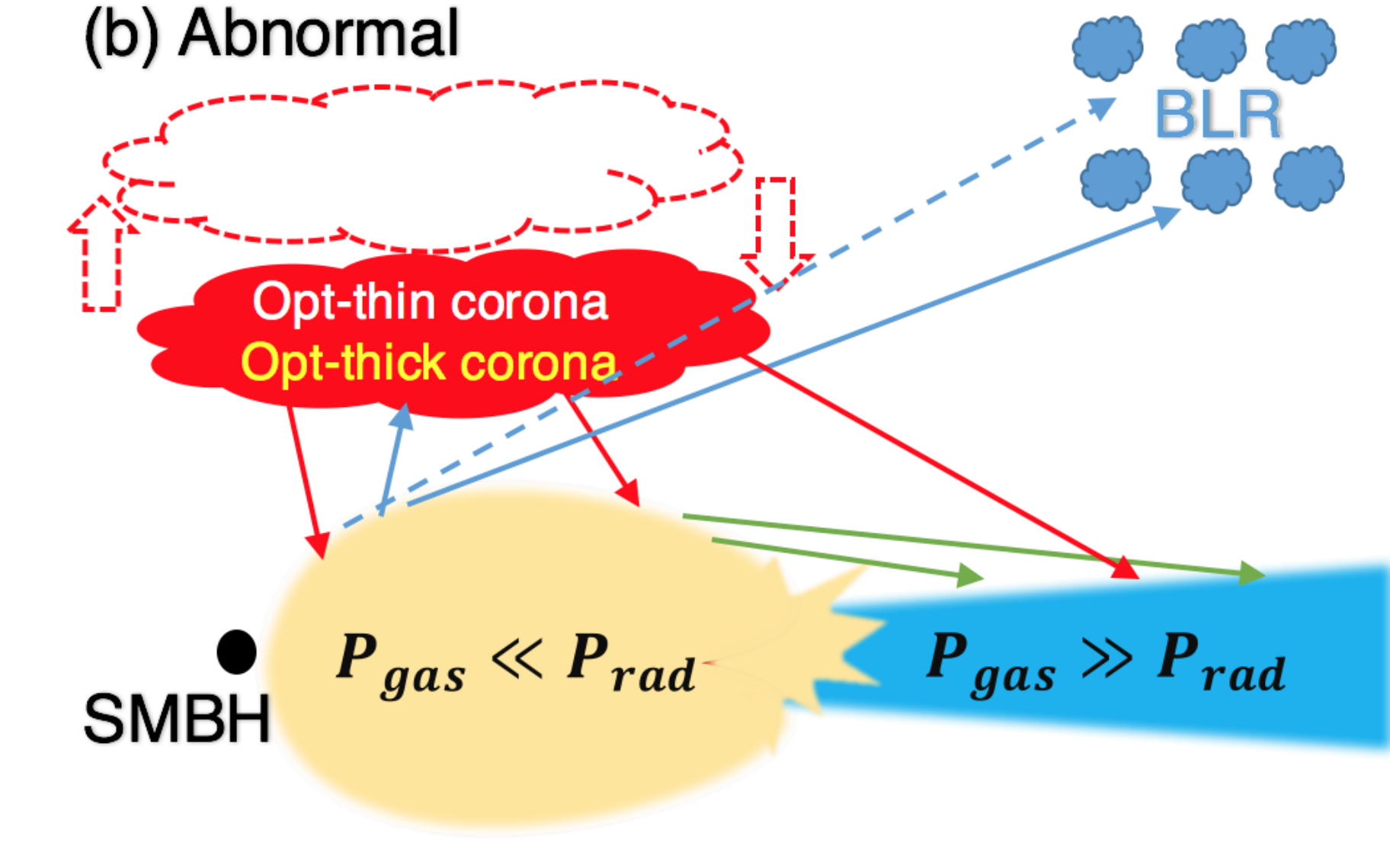}
\caption{An illustration of our simple model. The X-ray corona consists of an 
optically-thick layer and an optically-thin component. The former Comptonizes the seed EUV photons 
from the inner disk (i.e., the blue solid light ray) and is responsible for the soft X-ray excess; 
the later up-scatters the comptonized continuum and emits the power-law X-ray emission. The X-ray 
emission also illuminates the underlying accretion disk (i.e., the red light rays). As the 
X-ray corona moves towards/away 
the accretion disk, NGC 5548 enters the abnormal state. This is because the fraction of the 
UV/EUV-emitting regions that are covered by the corona increases with the distance of the corona to 
the disk decreasing. As the corona moves closer to the accretion disk, more UV/EUV photons will 
be absorbed by the corona and up-scattered to X-rays. As such, the X-ray spectrum would be softer 
and the coherence between X-ray and UV photons would be diluted. The ionizing EUV photons seen by 
the BLR gas are then depleted (see the dashed light ray), resulting in flux lost in BELs. High-ionization 
($E>50\ \mathrm{eV}$) BELs and high-ionizing continuum are closer to the SMBH than their low-ionization 
counterparts and are more easily covered by the corona. Hence, high-ionization BELs (e.g., CIV) suffer 
deficits by a larger fraction than the low-ionization BELs (e.g., H$\beta$). The line wings are more 
likely produced in the inner parts of the BLR. As a result, the line-wing fluxes are depleted by a 
larger factor. As NGC 5548 exits to the abnormal state, the corona rises up to the original position. 
Note that the UV/EUV emission regions are expected to be radiation pressure dominated (see text). 
}
\label{fig:toy}
\end{figure}

We propose a simple falling corona model that is illustrated in Figure~\ref{fig:toy} to explain the 
observational facts listed in Section~\ref{sect:obsf}. The corona is radially extended \citep[which 
is consistent with recent X-ray RM results; e.g.,][]{Wilkins2016} with a scale height equal to several 
times of the scale height of the accretion disk. The corona consists of two components, 
i.e., the upper optically thin and the lower optically thick layers. The former component produces 
power-law X-ray emission \citep[e.g.,][]{Haardt1993}; the later component is responsible for the soft 
X-ray excess \citep[e.g.,][]{Magdziarz1998}. The formation of the corona is not entirely clear. For 
simplicity, we assume that the corona itself dissipates energy due to the viscous accretion process. 
That is, a substantial accretion power is released in the corona. The radiative efficiency of the corona 
is less than that of the disk.\footnote{As demonstrated by \cite{MM1999} and \cite{Liu1999}, the corona 
and the hot accretion flow \citep[for a review, see][]{Yuan2014} are similar in many aspects.} Seed 
photons from the underlying thin disk \citep{SSD} are Compton up-scattered into X-ray photons. 

In the normal state, the corona is speculated to be in quasi-equilibrium with the disk, i.e., there is 
no significant spectral variations. As pointed out by \cite{Rozanska2015}, the optically 
thick (with optical depth $>5$) corona component cannot exist in hydrostatic equilibrium with the accretion 
disk. Let us image that, due to thermal instabilities in the corona, the cooling is more efficient than the 
heating, and the corona will condense and thus fall down. This non-equilibrium process is responsible 
for the anomalous behavior in NGC 5548. 

As the corona falls down, the covering fraction of the inner accretion disk increases. More 
seed photons emitted by the inner accretion disk will be inverse Compton up-scattered to 
X-rays. This leads to several observational consequences. First, spectral variations in the EUV 
to X-ray bands are expected, i.e., the spectrum will be softer. The variations of the normalization 
can lead to a positive inter-band correlation that drives the tight inter-band correlations in the 
normal state. On the contrary, the spectral variations can produce an anti-correlation between the seed 
and the up-scattered photons (e.g., EUV and X-ray). The observed variations in the X-ray 
or EUV bands are due to a combination of the variations of the normalization and spectral shape. 
As a result, the correlation between X-ray and EUV might be weak in the abnormal state. 
The UV variations could be driven by the X-ray and EUV variations. Hence, we expect weak correlation 
between EUV (BELs) and UV variations. Second, the fraction of EUV photons that ionize the BLR gas drops 
(e.g., the dashed light ray in Figure~\ref{fig:toy} is blocked) since the corona acts as a ``shield'' 
and up-scatters the EUV photons to X-ray ones. These two arguments can explain the observational facts 
$\# 1$ and $\# 4$. 

In our scenario, the depletion of BEL fluxes depends mainly on the reduction of EUV photons and the 
change of the fraction of EUV photons directly seen by the BLR gas. As the corona falls down, the decreasing 
amplitude of the fraction of the EUV fluxes directly seen by the BLR gas is smaller for clouds with smaller 
opening angles (i.e., the ratio of the height to the radius, $H/R$). There is evidence that 
the BLR is anisotropic with $H/R<1$ since the line profiles depend on inclination \citep[e.g.,][]{Krolik2001, 
Collin2006, Decarli2008, Runnoe2013, Shen2014}. The ratio $H/R$, as indicated by simple BLR kinematics 
modeling results \citep[e.g.,][]{Kollatschny2011}, decreases with $R$. On the other hand, in some failed 
dusty wind BLR models \citep[e.g.,][]{Baskin2017}, $H/R$ scales with the dust opacity; the dust opacity 
increases with frequency. As $R$ decreases, the accretion disk is hotter. That is, $H/R$ is expected 
to decrease with increasing $R$. Therefore, the EUV photons that ionize the inner parts (i.e., smaller radii)  
of the BLR gas will be preferentially scattered by the corona as it falls down. Since the line wings 
are more likely produced in these regions, the fraction of emission line lost in the wings is larger 
than that of the line center (i.e., the observational fact $\# 2$). For NGC 5548, the shape of the 
velocity-resolved time-lag profile is roughly symmetric about the line center \citep{Denney2009}. Therefore, 
the blue and red wings show similar anomalous behaviors.  

Compared with \Hbeta, high ionization emission lines (e.g., \CIV) respond to the variations 
of high energy ionizing continuum. The emission region of the high ionizing continuum is much more compact 
and closer to the SMBH than the low ionizing one. It is therefore possible that, as the corona falls 
down, the high ionizing continuum is more preferentially up-scattered into soft X-rays than the low ionizing 
one. Meanwhile, the \CIV\ gas is more closer to the SMBH than the \Hbeta\ gas. Hence, the fractions of 
depletion fluxes of high ionization emission lines are larger than those of low ionization emission lines. 
This can explain the observational fact $\# 3$. 

As NGC 5548 returns to the normal state, we speculate that the corona rises up to the 
original position. The transition timescale between the normal and abnormal states is controlled by 
the cooling, the falling, and the Compton timescales of the corona. The falling timescale is simply the 
dynamical timescale, which is smaller than the thermal timescale (see Eq.~\ref{eq:tcool}). The Compton 
timescale of the corona with the seed photons generated in the classical thin accretion disk is 
\citep{Ishibashi2012}
\begin{equation}
\label{eq:tct}
T_{\mathrm{ct}} = \frac{3m_{\mathrm{e}}c}{8\sigma_{\mathrm{TC}}\mu} = 
0.25\frac{0.1}{\eta}(\frac{R_{\mathrm{c}}}{100\ R_{\mathrm{S}}})^2\frac{0.1}{\dot{m}} 
\frac{M_{\mathrm{BH}}}{5\times 10^7\ M_{\odot}}\frac{0.1}{\alpha}\ \mathrm{days}
\end{equation}
where $m_{\mathrm{e}}$, $c$, $\sigma_{\mathrm{TC}}$, $\mu$, $R_{\mathrm{c}}$, $R_{\mathrm{S}}$, 
$\eta$ and $\dot{m}$ are the mass of electron, the speed of light, the Thomson cross section, the photon 
energy density, the radial distance to the SMBH, the Schwarzschild radius, the radiative efficiency, and 
the ratio of the accretion rate to the Eddington accretion rate (i.e., $1.3\times 10^{18}\ 
M_{\mathrm{BH}}/M_{\odot}\ \mathrm{g\ s^{-1}}$), respectively. 

For the corona with local viscous dissipation, the cooling time scale is \citep{Cao2016}
\begin{equation}
\label{eq:tcool}
T_{\mathrm{c}} = \frac{T_{\mathrm{th}}}{f}  = \frac{2\pi}{f\alpha \Omega_{\mathrm{K}}} 
= 80(\frac{R_{\mathrm{c}}}{100\ R_{\mathrm{S}}})^{\frac{3}{2}}\frac{M_{\mathrm{BH}}}{5\times 
10^7\ M_{\odot}}\frac{0.1}{\alpha f}\ \mathrm{days}
\end{equation}
where $T_{\mathrm{c}}$, $T_{\mathrm{th}}$, $\alpha$, and $\Omega_{\mathrm{K}}$ are the cooling 
timescale, the thermal timescale, the dimensionless viscosity, and Keplerian angular velocity, respectively. 
$f< 1$ is the ratio of the radiation to the gravitational energy of the corona. 
It is evident that the cooling timescale is much longer than the dynamical and the Compton timescales. 
Hence, the transition timescale is mainly controlled by the cooling timescale, which is $T_{\mathrm{c}} \sim 
50/2=25$ days (i.e., the observational fact \# 5; the factor of $2$ is introduced since NGC 5548 returns to 
the normal state). Therefore, the expected location of the corona is 
$R_{\mathrm{c}} < 40\ R_{\mathrm{S}}$ (i.e., Eq.~\ref{eq:tcool}), which is extended enough 
to obscure the EUV emission regions. Note that the obscured disk should be radiation-pressure dominated 
\citep{SSD}.

\section{The relation between the X-ray and UV variations}
\subsection{Revisiting the cross correlation between X-ray and UV emissions}
\label{sect:tlag}
Motivated by the observational fact $\# 4$ and our model, we expect significant spectral evolution from 
the normal to the abnormal state. As discussed in Section~\ref{sect:model}, the observed 
X-ray variability could be driven by both the variations in the normalization and spectral shape. Therefore, 
its power spectral density might change as NGC 5548 moves from the normal to the abnormal states. The ICCFs 
\citep[i.e., the interpolation cross-correlation function; see, e.g.,][]{Peterson1998} is depend on the 
power spectral density (PSD) 
of the light curves \citep{Welsh1999}. Meanwhile, it is also speculated that the variable BLR diffuse continuum 
emission can bias the time lag between the X-ray and UV variations. In the abnormal state, the BLR emission 
is suppressed. Therefore, the bias might be smaller than that of the normal state. Therefore, it is inappropriate 
to measure the time lag between X-ray and UV emission from the whole light curves. We therefore revisit the 
ICCF\footnote{We use PYCCF, Python Cross Correlation Function for reverberation mapping studies, to calculate 
the ICCFs. For details, see \url{http://ascl.net/code/v/1868}.} between the X-ray and the 
\textit{HST} $1367\ \rm{\AA}$ light curves. Following \cite{Edelson2015} and \cite{Fausnaugh2016}, we only 
interpolate the \textit{HST} $1367\ \rm{\AA}$ light curve and only consider the intensive monitoring period 
(i.e., from THJD$=6706$ to $6831$). We detrend the light curves by subtracting a $40$-day boxcar 
running mean. 

\begin{figure}
\epsscale{1.2}
\plotone{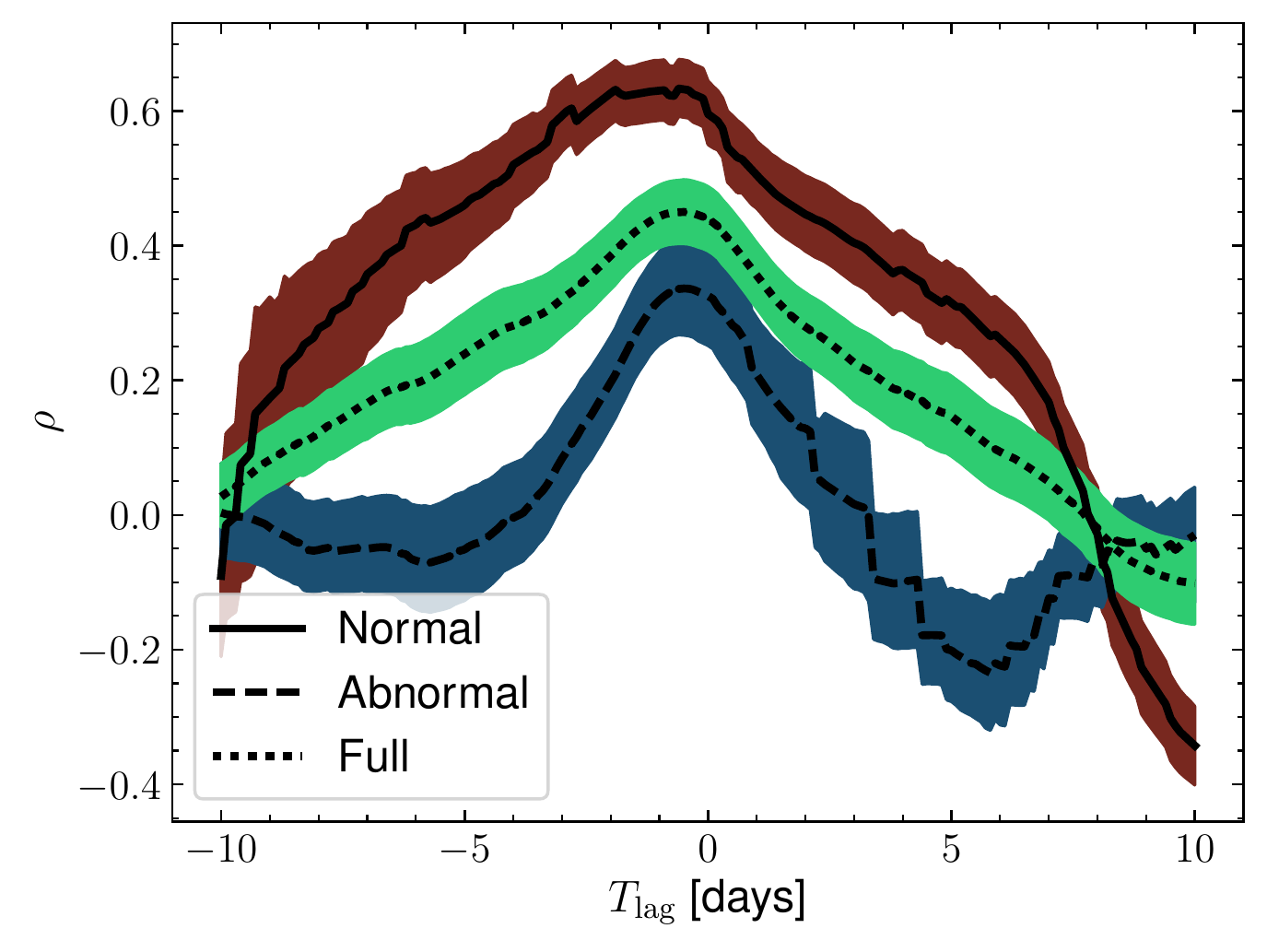}
\caption{The ICCFs between the hard X-ray and the \textit{HST} $1367\ \rm{\AA}$ light curves. The solid, 
dashed, and dotted curves represent the ICCFs for the first segment (i.e., the normal state), the second 
segment (i.e., the abnormal state), and the full light curves, respectively. The shaded regions indicate 
the corresponding $1\sigma$ uncertainties (via the FR/RSS simulation). It is evident that the hard X-ray 
and the \textit{HST} $1367\ \rm{\AA}$ variations are tightly correlated during the normal state. This 
correlation is weak during the abnormal state. In Figures ~\ref{fig:hx_ccf} \& 
~\ref{fig:hx_tlag}, we detrend the light curves by subtracting a $40$-day boxcar running mean. }
\label{fig:hx_ccf}
\end{figure}

We separately apply the ICCF to the first (i.e., the normal state) and second (i.e., the abnormal state) 
segments of the light curves. The epoch separating the two segments is THJD$=6747$ days \citep{Pei2017}. 

The ICCFs for the hard X-ray (i.e., $0.8$--$10$ keV) are shown in Figure~\ref{fig:hx_ccf}. The uncertainties 
are estimated by performing the bootstrapping simulations \citep[i.e., the FR/RSS; see][]{Peterson1998}. 
The correlation between the hard X-ray and the \textit{HST} $1367\ \rm{\AA}$ light curves is very tight in 
the normal state. On the contrary, the correlation is weak in the abnormal state. 

\begin{figure}
\epsscale{1.2}
\plotone{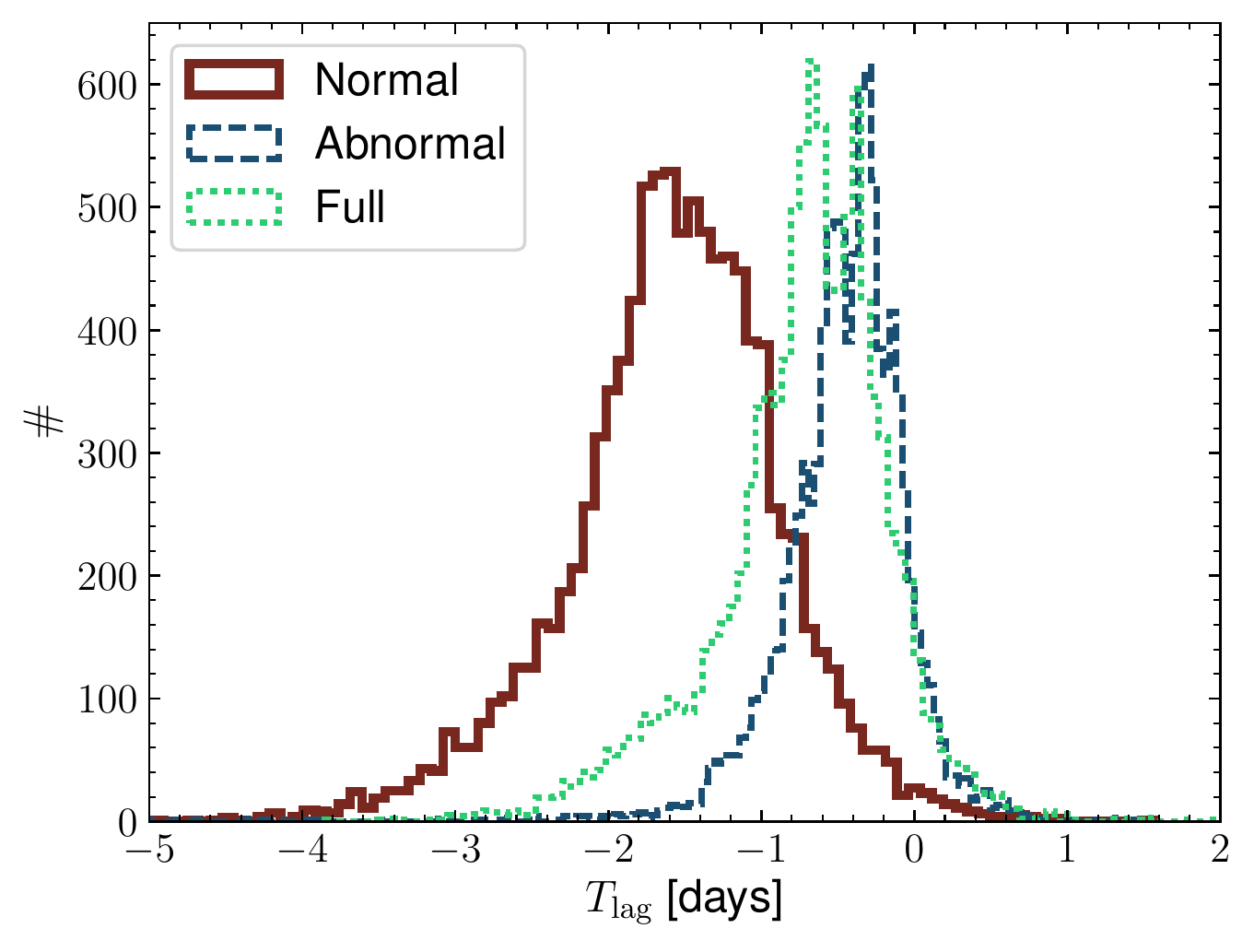}
\caption{The observed-frame hard X-ray time lag distributions from our FR/RSS simulations. The solid, 
dashed, and dotted histograms represent the distributions of $T_{\mathrm{lag}}$ for the first segment 
(i.e., the normal state), the second segment (i.e., the abnormal state), and the full light curves, 
respectively. Negative values indicate that the hard X-ray leads the \textit{HST} $1367\ \rm{\AA}$ 
variations. The time lag in the abnormal state is shorter than that of the normal state. 
Therefore, the time lag is significantly underestimated if the full light curves were adopted.
}
\label{fig:hx_tlag}
\end{figure}

We measure the centroids of the ICCFs in Figure~\ref{fig:hx_ccf}. The centroid is calculated by 
considering the mean (weighted by the correlation coefficient $\rho$) time lag for $\rho>0.8\ 
\rho_{\mathrm{max}}$. Figure~\ref{fig:hx_tlag} presents the distributions of the time lag of the 
hard X-ray with respect to the \textit{HST} $1367\ \rm{\AA}$. If we only adopt the first segment 
(i.e., in the normal state), the time lag is $-1.54^{-0.50}_{+0.44}$ days in the 
observed frame, where the uncertainties correspond to the $25$th and $75$th percentiles of the 
distribution. This result is significantly larger than the time lag measured from the full light 
curves ($-0.65^{-0.33}_{+0.28}$ days). Therefore, the time lag reported in 
\cite{Edelson2015} is significantly biased by the second segment of the light curves. 
Interestingly, but not surprisingly, the time lag of the abnormal state is 
$-0.4^{-0.25}_{+0.2}$ days, which is shorter than the normal state. 

\begin{figure}
\epsscale{1.2}
\plotone{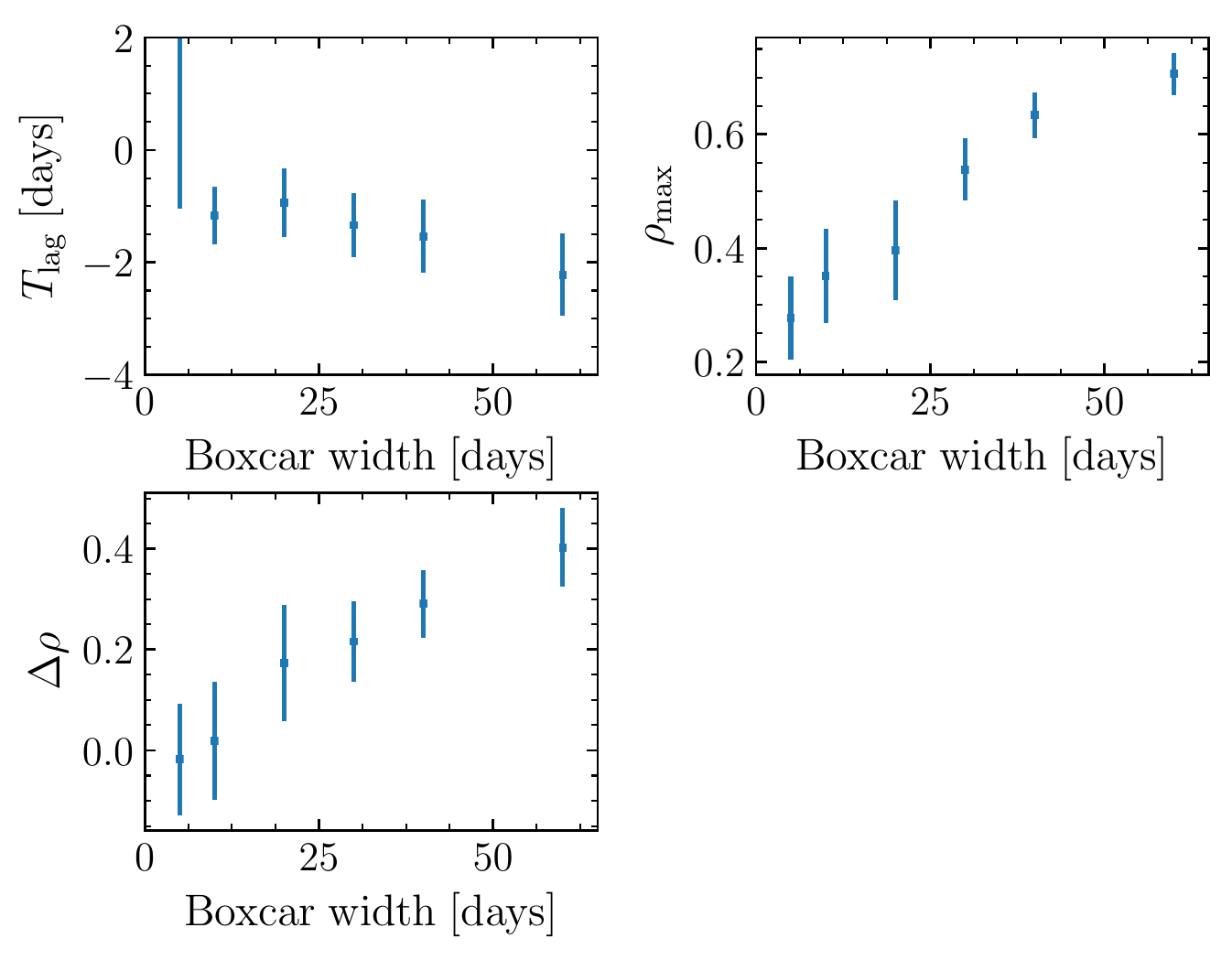}
\caption{The relation between the ICCF properties and the smoothing timescale for the 
normal state. Upper left: The time lag between the hard X-ray and the \textit{HST} $1367\ \rm{\AA}$ light 
curves as a function of the boxcar smoothing timescale. Upper right: The ICCF peak as a function of 
the boxcar smoothing timescale. Lower left: The difference between the positive ICCF 
peak and the negative one as a function of the boxcar smoothing timescale. The time lag is indeed 
smaller if we adopted shorter boxcar smoothing timescales. However, the correlation tends to be weaker; 
the difference between the positive and negative peaks is also diminished. }
\label{fig:detrend}
\end{figure}

These results would not be significantly changed if we measure the time lag by running \textit{JAVLIN} 
\citep{Zu2011}. However, the time lag depends on the detrending timescales. Motivated 
by \cite{McHardy2017}, we calculate the ICCFs and determine time lags for different detrending 
timescales, i.e., $5$ days, $10$ days, $20$ days, $30$ days, $40$ days, and $60$ days. The results 
for the normal state are presented in Figure~\ref{fig:detrend}. It is true that, for shorter smoothing 
timescales (e.g., $5$ or $10$ days), the time lag during the normal state tends to be smaller. However, 
the correlation between the hard X-ray and the \textit{HST} $1367\ \rm{\AA}$ light curve is also weaker. 
Moreover, the difference between the positive and negative peaks tend to diminish. Therefore, 
it is unclear whether the measured time lag for short smoothing timescales is robust. For the abnormal 
state, the ICCF properties (i.e., the time lag, the peak, and the difference between the positive and 
negative peaks) are largely independent from the smoothing timescale. If we consider the full light curves, 
the time lag and other ICCF properties approach those of the abnormal state. 

We also explore the ICCFs between the soft X-ray (i.e., $0.3$--$0.8$ keV) and the \textit{HST} $1367\ 
\rm{\AA}$ variations. We again find that the correlation of the normal state is significantly tighter 
than that of the abnormal state.

\subsection{Comparison with previous works}
The time lag between X-ray and UV emission is also explored in several other sources, 
e.g., NGC 2617 \citep{Shappee2014}, NGC 4151 \citep{Edelson2017}, NGC 4395 \citep{McHardy2016} and NGC 
4593 \citep{McHardy2017}. In these sources, the X-ray to UVW2 time lag is measured. 

We have also estimated the time lag between the hard X-ray and UVW2 emission for the normal, 
the abnormal, and the full light curves of NGC 5548. For the full light curves, the time lag is $-1.13^{-0.3}_{+0.25}$ 
days, which is perfectly consistent with that of \cite{Edelson2015}. However, if we divide the full 
light curve into two segments, the time lag changes. For the normal (abnormal) state, the time lag 
is $-3.08^{-0.48}_{+0.51}$ days ($-0.74^{-0.18}_{+0.22}$ days). Note that the time lag in the abnormal 
state agrees well with that of \cite{McHardy2014} who studied the \textit{Swift} observations prior to 
the AGN STORM project. Therefore, like other sources (except for NGC 4395), the time lag of NGC $5548$ in the 
normal state is vividly longer than the light-travel timescale of disk reprocessing (by a factor of $\sim 10$). 
However, the time lag of NGC $5548$ in the abnormal state again approaches the light-travel timescale albeit 
with a much weaker correlation.\footnote{We also explored the ICCFs between the hard X-ray and other UV-to-optical 
bands and obtained similar conclusions. The time lag in the normal state is much longer than the light-travel 
timescale of disk reprocessing; the time lag in the abnormal state approaches the light-travel timescale of 
disk reprocessing; the correlation in the abnormal state is much weaker than that of the normal state. Note 
that the properties of ICCFs between the \textit{HST} $1367\ \rm{\AA}$ and other UV-to-optical bands are not 
very sensitive to the anomalous behavior.}

\subsection{Implications to theoretical models}
\label{sect:implication}
\subsubsection{The implications of the time lag}
\label{sect:imp1}
We find that, after ignoring the second half of the AGN STORM light curves, the observed time 
lag between the hard X-ray and the \textit{HST} $1367\ \rm{\AA}$ emission is $\sim 1.5$ 
days (see Figure~\ref{fig:hx_tlag}). The $\sim 1.5$-day time lag is much longer than 
the light-travel timescale between the corona and the UV emission region ($\sim 0.2$ days if the 
distance is $d\sim 40\ R_{\mathrm{S}}$). However, the time lag for the abnormal state is roughly 
consistent with the light-travel timescale (i.e., within $\sim 1\ \sigma$ errorbar). 

The observed change of the time lags from the normal to the abnormal state does not 
necessarily imply a mighty change of the distance between the corona and the UV emission regions. 
It could also be caused by other factors, e.g., the variation of the transfer function \citep[i.e., 
the function that ``translates'' the X-ray light curves into UV light curves; see e.g., Eq.$\sim 2$ 
of][]{Horne2004} or the PSDs. If so, we should also observe the changes in both the X-ray and UV structure 
functions (i.e., the function that describes the variability as a function of the separating timescale). 
We adopted the following definition to calculate the structure function \citep[e.g.,][]{Sun2015},
\begin{equation}
\label{eq:iqr}
\mathrm{SF}(\Delta t) = \sqrt{(0.74\mathrm{IQR}(\Delta m))^2 - \widetilde{\sigma^2_{\rm{e}}}} \,
\end{equation}
where $\mathrm{IQR}(\Delta m)$ is the $25\%-75\%$ interquartile range of $\Delta m$ and 
$\widetilde{\sigma^2_{\rm{e}}}$ is the median of the measurement variance of $\Delta m$. $\Delta m=-2.5\log 
(f_2/f_1)$ is the difference between two observations (i.e., $f1$ and $f2$) separated by $\Delta t$. 
The constant $0.74$ normalizes the IQR to be equivalent to the standard deviation of a Gaussian distribution. 

\begin{figure}
\epsscale{2.4}
\plottwo{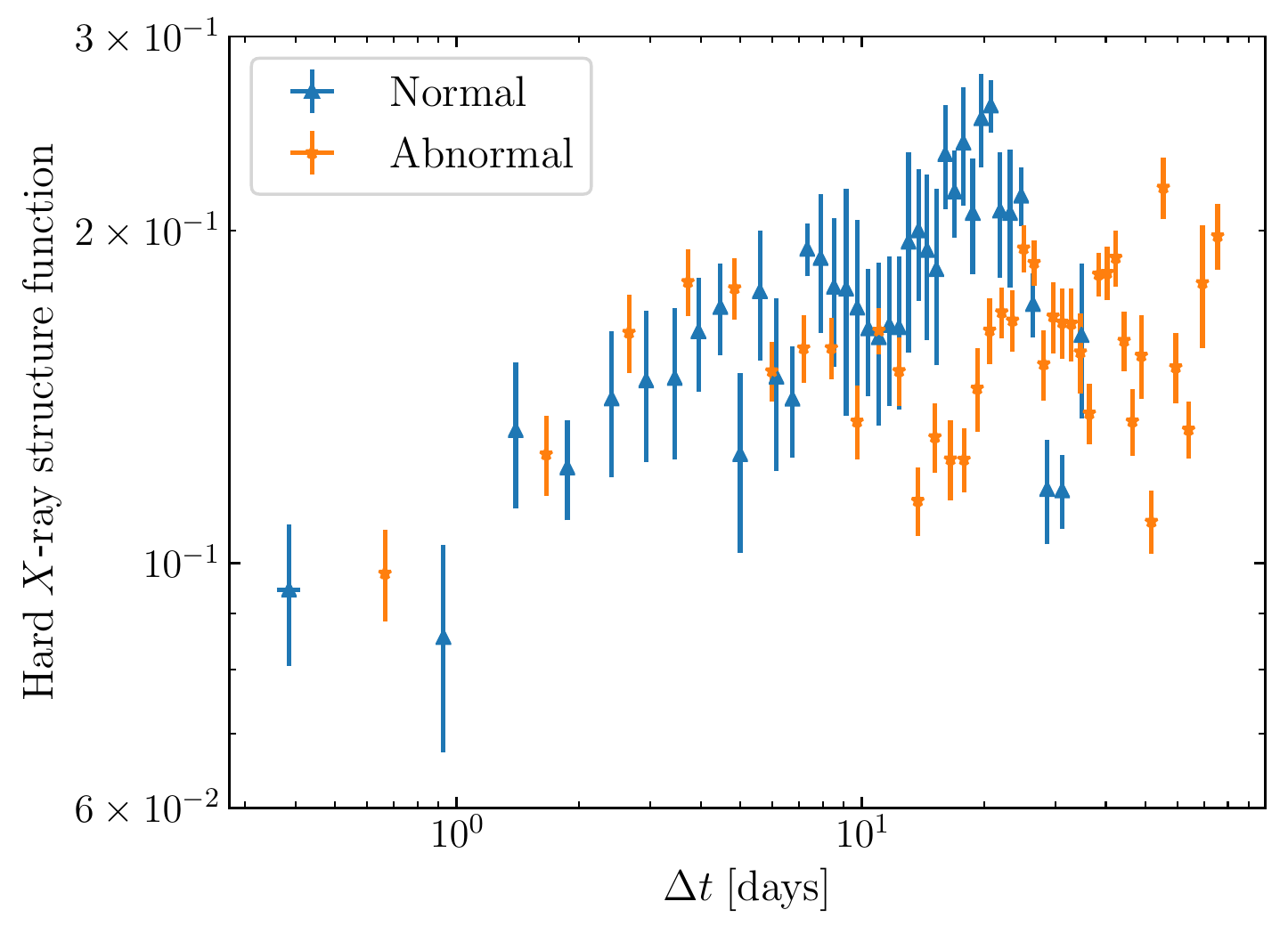}{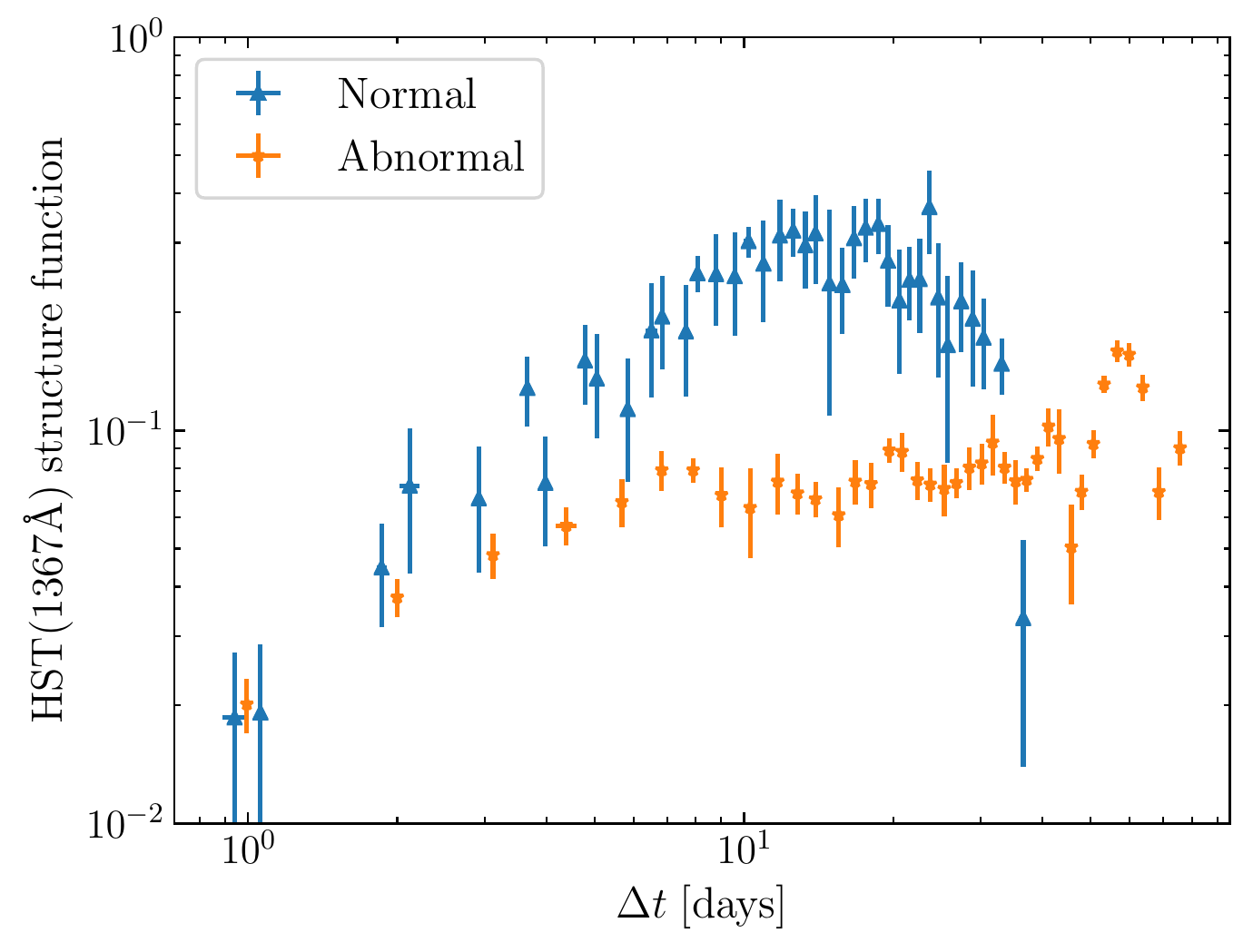}
\caption{The structure functions for the hard X-ray (upper) and the \textit{HST} $1367\ \rm{\AA}$ 
emission (lower). On timescales $\Delta t> 10$ days, NGC 5548 is evidently more variable in the normal state. 
During the abnormal state, the structure functions are more flat implying higher contributions of short 
timescale variations to the observed light curves. }
\label{fig:SF}
\end{figure}

The structure functions for the hard X-ray and the \textit{HST} $1367\ \rm{\AA}$ emission 
are presented in Figure~\ref{fig:SF}. On timescales $\Delta t> 10$ days, NGC 5548 is evidently less variable 
in the abnormal state. The evolution of the structure function in hard X-ray is not unexpected in our simple 
falling corona model. As mentioned in Section~\ref{sect:model}, both normalization and spectral variations 
are expected during the abnormal state. The two types of variability may cancel each other, resulting weaker 
observed variations. 

It is also interesting to mention that, for both two emission, the structure functions are 
more flat in the abnormal state, i.e., the contributions of short timescale (i.e., $< 10$ days) variations 
to the observed light curves are higher in the abnormal state than in the normal state. Therefore, our 
results appear to be consistent with the speculation that the time lag is consistent with the light-travel 
timescale of disk reprocessing if the long timescale variations are suppressed \citep[see also][]{McHardy2017}. 

How do we understand the time lags detected in the normal and abnormal states? There are several 
different scenarios. First, in addition to the accretion disk, there could be other reprocessors. For instance, 
as originally proposed by \cite{Korista2001} and recently suggested by \cite{Cackett2017} and \cite{McHardy2017}, 
the BLR gas might act as a second reprocessor. During the normal state, the contribution of the BLR component is 
strong; the observed $\sim 1.5$-day time lag reflects the response of the BLR reprocessor \citep[see, e.g., 
Figure~5 of][]{Korista2001}. As NGC 5548 enters the abnormal state, the contribution of the BLR gas is weaker; 
the time lag approaches the light-travel timescale between the corona and the underlying disk. 
This speculation is qualitatively consistent with the anomalous behavior of the BELs (i.e., the line emission 
is also ``lost'') and our model. Meanwhile, unlike the disk reprocessing, the BLR component should contribute 
more to relatively long timescale variations. Hence, the variations of the \textit{HST} $1367\ \rm{\AA}$ emission 
on $\Delta t> 10$ days are suppressed in the abnormal state. 

The second possibility is that the time lag is not related to the light-travel timescale. 
\cite{Cai2017} systematically explored the inter-band time lags by assuming a global common temperature 
fluctuation in the accretion disk. They further assumed that the timescale for the disk to respond to the 
global temperature fluctuations is radius-dependent (the timescale could be the dynamical or thermal timescale). 
They demonstrated that this phenomenological model has the potential to explain the observed time lags among 
UV-to-IR bands. If we assume that the corona varies in lockstep with the innermost (i.e., $<15\ R_{\mathrm{S}}$) 
accretion disk, we can also calculate the predicted time lag between the hard X-ray and the \textit{HST} 
$1367\ \rm{\AA}$ emission from the \cite{Cai2017} model. We find that the predicted time lag is 
$-2.69^{-0.73}_{+0.5}$ days that is roughly consistent with the time lag in the normal state. It is worth 
noting that, if the size of the corona increases, the difference between the timescale of the corona respond 
to the global temperature fluctuations and that of the UV emission decreases. As a result, the 
time lag can be smaller. 

The third possibility is that, as speculated by \cite{Gardner2017}, the dynamical or 
thermal timescales might be responsible for the observed inter-band time lags. We can then estimate the radius 
of the \textit{HST} $1367\ \rm{\AA}$ emission region by assuming that the observed $1.5$-day time 
lag is related to the dynamical timescale. We find that the required radius is $R_{\mathrm{rp}}\sim 40 
R_{\mathrm{S}}$ which is quite consistent with the expectation of the accretion theory \citep{Fausnaugh2016}. 
This scenario, however, has the difficulty to explain the short time lag in the abnormal state unless 
the dynamical or thermal timescale is somehow not important in the abnormal state.

\subsubsection{The implications of the correlation coefficient}
\label{sect:imp2}
\begin{figure}
\epsscale{1.2}
\plotone{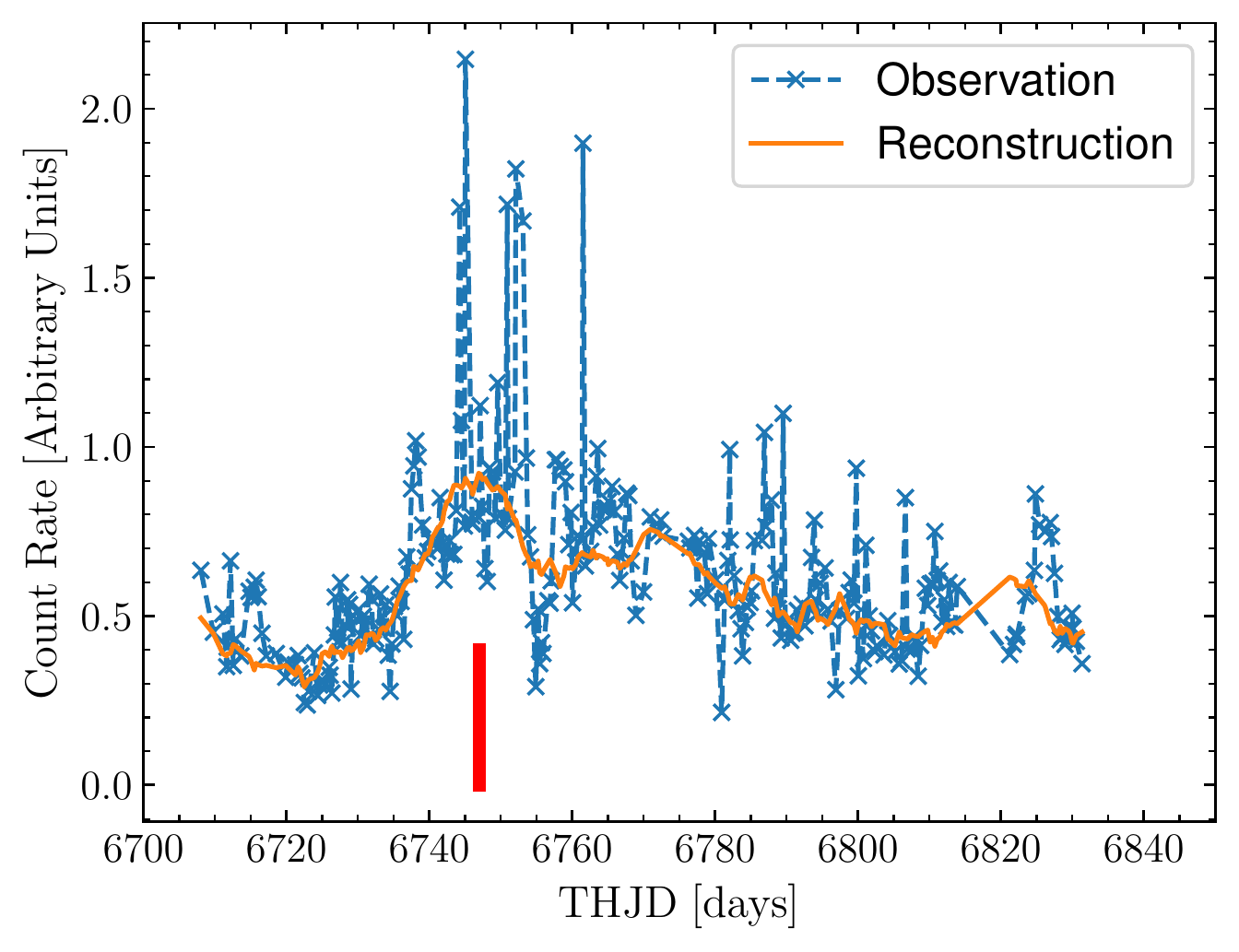}
\caption{A comparison between the observed (a $40$-day boxcar running mean is subtracted) and reconstructed 
hard X-ray light curves. For THJD$<6747$ (i.e., the ``normal'' state), the two light curves are well matched. 
A similar correlation is not observed in the ``abnormal'' state. The red thick vertical line indicates THJD$= 
6747$ days. }
\label{fig:hx_lc}
\end{figure}

As we mentioned in Section~\ref{sect:tlag} (and Figure~\ref{fig:hx_ccf}), the correlation between 
the hard X-ray and the UV emission is rather tight if we ignore the second half of the light curves (i.e., the 
abnormal light curves). Indeed, the maximum correlation coefficient $\rho_{\mathrm{max}}=0.62\pm 0.04$. We 
also use the lag-corrected \textit{HST} $1367\ \rm{\AA}$ light curve to reconstruct the associated X-ray emission. 
First, we fitted the \textit{continuous time first-order autoregressive process} \citep[i.e., CAR(1); see, 
e.g.,][]{Kelly2009} to the lag-corrected \textit{HST} $1367\ \rm{\AA}$ light curve (a $40$-day boxcar running mean 
is subtracted) using \textit{CARMA}\footnote{This package can be downloaded from 
\url{https://github.com/brandonckelly/carma_pack}.} \citep{Kelly2014}. 
Second, we adopted the best-fitting CAR(1) model to simulate the \textit{HST} $1367\ \rm{\AA}$ flux associated with 
the hard X-ray flux (a $40$-day boxcar running mean is also subtracted) at each epoch. Third, we fitted 
a linear relation to the first half (i.e., THJD$<6747$ days) of two light curves. Each data point is weighted 
by its measurement error. Using the best-fitting linear relation and the simulated \textit{HST} $1367\ 
\rm{\AA}$ flux, we reconstructed the hard X-ray flux. In Figure~\ref{fig:hx_lc}, we present both the observed 
and reconstructed hard X-ray light curves. For THJD$<6747$ days, the two light curves are reasonably matched. 
The tight correlation between the X-ray and UV emission is also observed for another source, NGC 4593 
\citep{McHardy2017}. 

During the abnormal state, the correlation between the hard X-ray and the \textit{HST} $1367\ 
\rm{\AA}$ emission is rather poor ($\rho_{\rm{max}}=0.34$). There are ``outliers'' in the hard X-ray light curve, 
i.e., the count rate increases by a factor of $> 2$ within a few days. Similar features are not observed in 
the \textit{HST} $1367\ \rm{\AA}$ light curve. However, it should be noted that the poor correlation is not 
caused by these outliers. We tried to remove these X-ray outliers and recalculated the ICCF of the abnormal 
state. The maximum correlation coefficient $\rho_{\rm{max}}=0.36$, i.e., the correlation is still poor. 

How do we understand the change of the correlation coefficient? For instance, according to our falling 
corona model, the UV variations can be driven by both the X-ray fluctuations and the changes of the solid angle subtended 
by the UV emission regions from the corona. Meanwhile, the time lag of the abnormal state is small. The ICCF depends more 
critically on the interpolation since the sampling interval is larger than the observed time lag. Therefore, the correlation 
between the X-ray and UV variations is diluted. 

\cite{Gardner2017} explored the expected correlation between the hard X-ray and the UV emission from 
the X-ray reprocessing scenario. They found that the simulated correlation is much tighter than the AGN STORM light 
curves and concluded that the observed UV-optical variations are unlikely driven by the X-ray variability. Instead, 
they proposed that the hot X-ray corona cannot directly illuminate the accretion disk that is shielded by a UV torus. 
The X-ray emission heats the UV torus; the UV torus then illuminates the underlying accretion disk. As X-ray emission 
varies, a heating wave dissipates outwards, which drives the variability of the disk emission (e.g., optical emission). 
That is, the variability of long-wavelength (i.e., optical, IR) emission is controlled by the UV torus rather than 
the corona. Therefore, the correlation between the hard X-ray and UV emission can be poor. This model seems to be able 
to explain the poor correlation in the abnormal state. However, in the normal state, $\rho_{\mathrm{max}}= 
0.62$ (if we interpolate the hard X-ray light curve, then $\rho_{\mathrm{max}}=0.81\pm 0.05$) is not 
radically smaller than the expectation of the simple X-ray reprocessing \cite[the simulated $\rho_{\mathrm{max}}\approxeq 
0.8\sim 0.9$; see Figures~4 \& 5 of][]{Gardner2017}. Hence, the UV torus appears to be weak or absent in the normal state. 

The weak correlation between the X-ray and the UV variations is also observed in another source NGC 4151 
\citep{Edelson2017}. However, it is quite possible that the observed X-ray emission does not vary in phase with the one 
that illuminates the accretion disk. For instance, gas clouds can move into/away from light of sight on short timescales 
\citep[for NGC 4151, see][]{Wang2010}. Therefore, the nature of the weak correlation could be different from that of 
the abnormal state of NGC 5548.

\section{Summary}
\label{sect:summ}
We propose a simple falling corona model to explain the anomalous behavior of the broad emission lines 
in NGC 5548. In our model, NGC 5548 enters the abnormal state as the corona falls down towards the accretion 
disk. During this process, the covering factor of the corona increases. This process can naturally explain 
the observational facts summarized in Section~\ref{sect:obsf}. 

We demonstrate that the time lag between X-ray and UV emission reported in the previous work is 
biased due to the anomalous behavior. The time lag in the normal state is $-1.54^{-0.50}_{+0.44}$ 
days (in the observed frame). The light-travel time delay cannot account for the $\sim 1.5$-day 
time lag. As NGC 5548 enters the abnormal state, the time lag approaches the light-travel 
timescale between the corona and the accretion disk. We speculate that the time lag in the normal state is 
related to reprocessing from the BLR gas. As NGC 5548 enters the abnormal state, the BLR contribution is 
smaller. Therefore, the time lag reflects disk reprocessing. However, other possibilities can not be excluded. 

We also show that the correlation between the hard X-ray and UV variations can be tight 
($\rho_{\mathrm{max}}\sim 0.6$) if the anomalous behavior is properly excluded. The correlation coefficient 
is roughly consistent with the simple X-ray reprocessing scenario. Therefore, the UV torus component proposed 
by \cite{Gardner2017} is weak or absent in the normal state of NGC 5548. 

Our model makes a clear prediction. Instead of falling down, the corona can also rise up to a larger scale 
of height. During the rising process, another type of the anomalous behavior of the BELs occurs. That is, 
while the correlation between the variations of quasar continua and BELs is destroyed, the BELs also show 
excess fluxes (i.e., $f_{\mathrm{lost}} <0$) than the reconstructed BEL light curves (e.g., from the 
\textit{HST} $1367\ \rm{\AA}$ light curve) and the X-ray spectrum is harder. Future RM experiments can 
verify such a scenario. 

\acknowledgments
We acknowledge Prof. Wei-Min Gu and Prof. Jun-Xian Wang for beneficial discussion. M.Y.S. and Y.Q.X. acknowledge 
the support from NSFC-11603022, NSFC-11473026, NSFC-11421303, the 973 Program (2015CB857004), the 
China Postdoctoral Science Foundation (2016M600485), the CAS Frontier Science Key Research Program 
(QYZDJ-SSW-SLH006), and the Fundamental Research Funds for the Central Universities. Z.Y.C. 
acknowledges the support from NSFC-11503024. 

\software{Astropy \citep{Astropy2013}, CARMA \citep{Kelly2014}, Matplotlib \citep{Hunter2007}, 
Numpy \& Scipy \citep{scipy}}, PYCCF \citep{Sun2018}


\begin{thebibliography}{}
\bibitem[Astropy Collaboration et al.(2013)]{Astropy2013} Astropy Collaboration, Robitaille, T.~P., Tollerud, E.~J., et al.\ 2013, \aap, 558, A33 
\bibitem[Baskin \& Laor(2017)]{Baskin2017} Baskin, A., \& Laor, A.\ 2017, arXiv:1711.00025
\bibitem[Bentz et al.(2010a)]{Bentz2010a} Bentz, M.~C., Horne, K., Barth, A.~J., et al.\ 2010a, \apjl, 720, L46 
\bibitem[Bentz \& Katz(2015)]{Bentz2015} Bentz, M.~C., \& Katz, S.\ 2015, \pasp, 127, 67 
\bibitem[Bentz et al.(2010b)]{Bentz2010b} Bentz, M.~C., Walsh, J.~L., Barth, A.~J., et al.\ 2010b, \apj, 716, 993 
\bibitem[Blandford \& McKee(1982)]{BM1982} Blandford, R.~D., \& McKee, C.~F.\ 1982, \apj, 255, 419 
\bibitem[Cackett et al.(2017)]{Cackett2017} Cackett, E.~M., Chiang, C.-Y., McHardy, I., et al.\ 2017, arXiv:1712.04025
\bibitem[Cai et al.(2017)]{Cai2017} Cai, Z.-Y., Wang, J.-X., Zhu, F.-F., et al.\ 2017, arXiv:1711.06266
\bibitem[Cao(2016)]{Cao2016} Cao, X.\ 2016, \apj, 817, 71
\bibitem[Collier et al.(1998)]{Collier1998} Collier, S.~J., Horne, K., Kaspi, S., et al.\ 1998, \apj, 500, 162
\bibitem[Collin et al.(2006)]{Collin2006} Collin, S., Kawaguchi, T., Peterson, B.~M., \& Vestergaard, M.\ 2006, \aap, 456, 75
\bibitem[Decarli et al.(2008)]{Decarli2008} Decarli, R., Labita, M., Treves, A., \& Falomo, R.\ 2008, \mnras, 387, 1237
\bibitem[Denney et al.(2009)]{Denney2009} Denney, K.~D., Peterson, B.~M., Pogge, R.~W., et al.\ 2009, \apjl, 704, L80
\bibitem[Denney et al.(2010)]{Denney2010} Denney, K.~D., Peterson, B.~M., Pogge, R.~W., et al.\ 2010, \apj, 721, 715
\bibitem[De Rosa et al.(2015)]{Derosa2015} De Rosa, G., Peterson, B.~M., Ely, J., et al.\ 2015, \apj, 806, 128  
\bibitem[Du et al.(2014)]{Du2014} Du, P., Hu, C., Lu, K.-X., et al.\ 2014, \apj, 782, 45 
\bibitem[Edelson et al.(1996)]{Edelson1996} Edelson, R.~A., Alexander, T., Crenshaw, D.~M., et al.\ 1996, \apj, 470, 364
\bibitem[Edelson et al.(2017)]{Edelson2017} Edelson, R., Gelbord, J., Cackett, E., et al.\ 2017, \apj, 840, 41 
\bibitem[Edelson et al.(2015)]{Edelson2015} Edelson, R., Gelbord, J.~M., Horne, K., et al.\ 2015, \apj, 806, 129 
\bibitem[Fausnaugh et al.(2016)]{Fausnaugh2016} Fausnaugh, M.~M., Denney, K.~D., Barth, A.~J., et al.\ 2016, \apj, 821, 56 
\bibitem[Gardner \& Done(2017)]{Gardner2017} Gardner, E., \& Done, C.\ 2017, \mnras, 470, 3591 
\bibitem[Goad et al.(2016)]{Goad2016} Goad, M.~R., Korista, K.~T., De Rosa, G., et al.\ 2016, \apj, 824, 11
\bibitem[Grier et al.(2012)]{Grier2012} Grier, C.~J., Peterson, B.~M., Pogge, R.~W., et al.\ 2012, \apj, 755, 60 
\bibitem[Grier et al.(2017a)]{Grier2017a} Grier, C.~J., Pancoast, A., Barth, A.~J., et al.\ 2017a, \apj, 849, 146 
\bibitem[Grier et al.(2017b)]{Grier2017b} Grier, C.~J., Trump, J.~R., Shen, Y., et al.\ 2017b, \apj, 851, 21 
\bibitem[Haardt \& Maraschi(1993)]{Haardt1993} Haardt, F., \& Maraschi, L.\ 1993, \apj, 413, 507
\bibitem[Horne et al.(2004)]{Horne2004} Horne, K., Peterson, B.~M., Collier, S.~J., \& Netzer, H.\ 2004, \pasp, 116, 465
\bibitem[Hunter(2007)]{Hunter2007} Hunter, J.~D.\ 2007, Computing in Science and Engineering, 9, 90 
\bibitem[Ishibashi \& Courvoisier(2012)]{Ishibashi2012} Ishibashi, W., \& Courvoisier, T.~J.-L.\ 2012, \aap, 540, L2 
\bibitem[Kaspi et al.(2000)]{Kaspi2000} Kaspi, S., Smith, P.~S., Netzer, H., et al.\ 2000, \apj, 533, 631  
\bibitem[Kaspi et al.(2007)]{Kaspi2007} Kaspi, S., Brandt, W.~N., Maoz, D., et al.\ 2007, \apj, 659, 997 
\bibitem[Kelly et al.(2009)]{Kelly2009} Kelly, B.~C., Bechtold, J., \& Siemiginowska, A.\ 2009, \apj, 698, 895-910 
\bibitem[Kelly et al.(2014)]{Kelly2014} Kelly, B.~C., Becker, A.~C., Sobolewska, M., Siemiginowska, A., \& Uttley, P.\ 2014, \apj, 788, 33
\bibitem[Korista \& Goad(2001)]{Korista2001} Korista, K.~T., \& Goad, M.~R.\ 2001, \apj, 553, 695
\bibitem[Krolik et al.(1991)]{Krolik1991} Krolik, J.~H., Horne, K., Kallman, T.~R., et al.\ 1991, \apj, 371, 541
\bibitem[Krolik(2001)]{Krolik2001} Krolik, J.~H.\ 2001, \apj, 551, 72
\bibitem[Kollatschny \& Zetzl(2011)]{Kollatschny2011} Kollatschny, W., \& Zetzl, M.\ 2011, \nat, 470, 366  
\bibitem[Liu et al.(1999)]{Liu1999} Liu, B.~F., Yuan, W., Meyer, F., Meyer-Hofmeister, E., \& Xie, G.~Z.\ 1999, \apjl, 527, L17
\bibitem[Magdziarz et al.(1998)]{Magdziarz1998} Magdziarz, P., Blaes, O.~M., Zdziarski, A.~A., Johnson, W.~N., \& Smith, D.~A.\ 1998, \mnras, 301, 179 
\bibitem[Mathur et al.(2017)]{Mathur2017} Mathur, S., Gupta, A., Page, K., et al.\ 2017, \apj, 846, 55 
\bibitem[Meyer-Hofmeister \& Meyer(1999)]{MM1999} Meyer-Hofmeister, E., \& Meyer, F.\ 1999, \aap, 348, 154
\bibitem[McHardy et al.(2014)]{McHardy2014} McHardy, I.~M., Cameron, D.~T., Dwelly, T., et al.\ 2014, \mnras, 444, 1469
\bibitem[McHardy et al.(2017)]{McHardy2017} McHardy, I., Connolly, S., Cackett, K.~E., et al.\ 2017, arXiv:1712.04852
\bibitem[McHardy et al.(2016)]{McHardy2016} McHardy, I.~M., Connolly, S.~D., Peterson, B.~M., et al.\ 2016, Astronomische Nachrichten, 337, 500 
\bibitem[Onken et al.(2004)]{Onken2004} Onken, C.~A., Ferrarese, L., Merritt, D., et al.\ 2004, \apj, 615, 645 
\bibitem[Pal \& Naik(2018)]{Pal2018} Pal, M., \& Naik, S.\ 2018, \mnras, 474, 5351
\bibitem[Pei et al.(2017)]{Pei2017} Pei, L., Fausnaugh, M.~M., Barth, A.~J., et al.\ 2017, \apj, 837, 131 
\bibitem[Peterson et al.(2002)]{Peterson2002} Peterson, B.~M., Berlind, P., Bertram, R., et al.\ 2002, \apj, 581, 197 
\bibitem[Peterson et al.(1998)]{Peterson1998} Peterson, B.~M., Wanders, I., Horne, K., et al.\ 1998, \pasp, 110, 660 
\bibitem[Peterson(2014)]{Peterson2014} Peterson, B.~M.\ 2014, \ssr, 183, 253 
\bibitem[Reynolds \& Nowak(2003)]{Reynolds2003} Reynolds, C.~S., \& Nowak, M.~A.\ 2003, \physrep, 377, 389 
\bibitem[R{\'o}{\.z}a{\'n}ska et al.(2015)]{Rozanska2015} R{\'o}{\.z}a{\'n}ska, A., Malzac, J., Belmont, R., Czerny, B., \& Petrucci, P.-O.\ 2015, \aap, 580, A77
\bibitem[Runnoe et al.(2013)]{Runnoe2013} Runnoe, J.~C., Brotherton, M.~S., Shang, Z., Wills, B.~J., \& DiPompeo, M.~A.\ 2013, \mnras, 429, 135 
\bibitem[Sergeev et al.(2005)]{Sergeev2005} Sergeev, S.~G., Doroshenko, V.~T., Golubinskiy, Y.~V., Merkulova, N.~I., \& Sergeeva, E.~A.\ 2005, \apj, 622, 129 
\bibitem[Shakura \& Sunyaev(1973)]{SSD} Shakura, N.~I., \& Sunyaev, R.~A.\ 1973, \aap, 24, 337
\bibitem[Shappee et al.(2014)]{Shappee2014} Shappee, B.~J., Prieto, J.~L., Grupe, D., et al.\ 2014, \apj, 788, 48
\bibitem[Shen \& Ho(2014)]{Shen2014} Shen, Y., \& Ho, L.~C.\ 2014, \nat, 513, 210 
\bibitem[Shen et al.(2016)]{Shen2016} Shen, Y., Horne, K., Grier, C.~J., et al.\ 2016, \apj, 818, 30 
\bibitem[Starkey et al.(2017)]{Starkey2017} Starkey, D., Horne, K., Fausnaugh, M.~M., et al.\ 2017, \apj, 835, 65
\bibitem[Starkey et al.(2016)]{Starkey2016} Starkey, D.~A., Horne, K., \& Villforth, C.\ 2016, \mnras, 456, 1960
\bibitem[Sun et al.(2018)]{Sun2018} Sun, M., Grier, C.~J., Peterson, B.~M.\ 2018, PYCCF, \url{https://ascl.net/code/v/1868}
\bibitem[Sun et al.(2015)]{Sun2015} Sun, M., Trump, J.~R., Shen, Y., et al.\ 2015, \apj, 811, 42
\bibitem[Van Der Walt et al.(2011)]{scipy} Van Der Walt, S., Colbert, S.~C., \& Varoquaux, G.\ 2011, arXiv:1102.1523 
\bibitem[Vestergaard \& Peterson(2006)]{Vestergaard2006} Vestergaard, M., \& Peterson, B.~M.\ 2006, \apj, 641, 689 
\bibitem[Wanders et al.(1997)]{Wanders1997} Wanders, I., Peterson, B.~M., Alloin, D., et al.\ 1997, \apjs, 113, 69
\bibitem[Wang et al.(2010)]{Wang2010} Wang, J., Risaliti, G., Fabbiano, G., et al.\ 2010, \apj, 714, 1497
\bibitem[Welsh(1999)]{Welsh1999} Welsh, W.~F.\ 1999, \pasp, 111, 1347
\bibitem[Wilkins et al.(2016)]{Wilkins2016} Wilkins, D.~R., Cackett, E.~M., Fabian, A.~C., \& Reynolds, C.~S.\ 2016, \mnras, 458, 200 
\bibitem[Yuan \& Narayan(2014)]{Yuan2014} Yuan, F., \& Narayan, R.\ 2014, \araa, 52, 529
\bibitem[Zu et al.(2011)]{Zu2011} Zu, Y., Kochanek, C.~S., \& Peterson, B.~M.\ 2011, \apj, 735, 80
\end{thebibliography}
\end{document}